\documentclass[aps,prb,twocolumn,showpacs,superscriptaddress]{revtex4}

\usepackage{graphicx}
\usepackage{amsmath}
\graphicspath{{fig/}{figsgaoerb/}}
\usepackage[colorlinks,linkcolor=blue,anchorcolor=blue,citecolor=blue]{hyperref}

\begin{document}
\title{Dilution induced magnetic localization in Rb(Co$_{1-x}$Ni$_{x}$)$_{2}$Se$_{2}$ single crystals }	
\author{Hui Liu}
\author{Mengwu Huo}
\author{Chaoxin Huang}
\author{Xing Huang}
\author{Hualei Sun}
\author{Lan Chen}
\affiliation{Center for Neutron Science and Technology, Guangdong Provincial Key Laboratory of Magnetoelectric Physics and Devices, School of Physics, Sun Yat-Sen University, Guangzhou, 510275, China }
\author{Juping Xu}
\affiliation{Institute of High Energy Physics, Chinese Academy of Sciences, Beijing 100049, China}
\affiliation{Spallation Neutron Source Science Center, Dongguan 523803, China}
\author{Wen Yin}
\affiliation{Institute of High Energy Physics, Chinese Academy of Sciences, Beijing 100049, China}
\affiliation{Spallation Neutron Source Science Center, Dongguan 523803, China}
\author{Runxia Li}
\email{5327368@qq.com}
\affiliation{Materials Science and Engineering, Dongguan University of Technology, Dongguan 523808, China}
\author{Meng Wang}
\email{wangmeng5@mail.sysu.edu.cn}
\affiliation{Center for Neutron Science and Technology, Guangdong Provincial Key Laboratory of Magnetoelectric Physics and Devices, School of Physics, Sun Yat-Sen University, Guangzhou, 510275, China }
\begin{abstract}

We report experimental studies on a series of Rb(Co$_{1-x}$Ni$_{x}$)$_{2}$Se$_{2}$ (0.02 $\leq x \leq $ 0.9) powder and single crystal samples using x-ray diffraction, neutron diffraction, magnetic susceptibility, and electronic transport measurements. All compositions are metallic and adopt the body-centered tetragonal structure with $I4/mmm$ space group. 
Anisotropic magnetic susceptibilities measured on single crystal samples suggest that Rb(Co$_{1-x}$Ni$_{x}$)$_{2}$Se$_{2}$ undergo an evolution from ferromagnetism to antiferromagnetism, and finally to paramagnetism with increasing Ni concentration.
Neutron diffraction measurements on the samples with $x$ = 0.1, 0.4, and 0.6 reveal an $A$-type antiferromagnetic order with moments lying in the $ab$ plane. The moment size changes from 0.69 ($x=0.1$) to 2.80$\mu_B$ ($x=0.6$) per Co ions. Our results demonstrate that dilution of the magnetic Co ions by substitution of nonmagnetic Ni ions induces magnetic localization and evolution from itinerant to localized magnetism in Rb(Co$_{1-x}$Ni$_{x}$)$_{2}$Se$_{2}$.	
\end{abstract}

\maketitle

\section{INTRODUCTION}	

Layered transition metal chalcogenides with the ThCr$_{2}$Si$_{2}$-type structure show versatile characteristics, such as superconductivity, heavy fermion behavior, and various magnetic orders due to the sensitivity of the magnetic and electronic correlations to the atomic distances and compositions.\cite{Dai2015b,Gegenwart2008,Berger2002,Metals1989,Ronneteg2010,Narasimhan1975,Ying2012a,Yang2013} 
The iron chalcogenide compounds $A_{x}$Fe$_{2}$Se$_{2}$ ($A=$ K, Rb, Cs, and Tl) exhibit superconductivity with a transition temperature at $\sim$30 K.\cite{Guo2010,Fang2010} While by inducing iron vacancies, for compositions with $\delta=0.4$ and 0.5, $A_{x}$Fe$_{2-\delta}$Se$_{2}$ become insulating and show iron vacancy and antiferromagnetic (AFM) orders.\cite{Bao2011,Zhao2012,Wang2016} In the insulating states, the Fe$^{2+}$ with fully localized electrons are in a high spin state.
The Co-doping in $A_{x}$Fe$_{2}$Se$_{2}$ suppresses the superconductivity rapidly and induces metallic magnetic ground states.\cite{Huang2021,Lizarraga2004,Liu2018,Yang2013,Ryu2015a}

By replacing Fe for Ni, the $A$Ni$_{2}$Se$_{2}$ compounds show Pauli paramagnetic (PM), multi-band characteristics, and superconductivity with superconducting transition temperatures at $0.8\sim3.7$ K.\cite{Lei2014,Chen2016,Wang2013c,Ryu2015,Liu2022}
While the isostructural compounds KCo$_{2}$Se$_{2}$ and RbCo$_{2}$Se$_{2}$ exhibit ferromagnetic (FM) order with Curie temperatures of $T_{\text{C}}=74$ and 83 K, respectively.\cite{Huang2021,Yang2013} CsCo$_{2}$Se$_{2}$ is an $A$-type AFM metal with moments lying in the $ab$ plane.\cite{VonRohr2016,Liu2018} The size of the in-plane ordered moments of Co ions in RbCo$_{2}$Se$_{2}$ is $\sim$0.60$\mu_{B}$ determined from neutron diffraction measurement, indicating an itinerant origin.\cite{Huang2021} As a simple expectation, Ni substitution on the Co sites would progressively tune the ground states of Rb(Co$_{1-x}$Ni$_{x}$)$_{2}$Se$_{2}$ from FM order to PM state. However, in previous studies of Tl(Co$_{1-x}$Ni$_{x}$)$_{2}$Se$_{2}$ and Tl(Co$_{1-x}$Ni$_{x}$)$_{2}$S$_{2}$, the AFM order was found to exist in a large region of the Ni doping level. Although the magnetic order of TlCo$_{2}$Se$_{2}$ is AFM and that of TlCo$_{2}$S$_{2}$ is FM.\cite{Newmark1989,Huan1989} 
These results indicate that the effect of Ni substitution on the magnetism in $A$(Co$_{1-x}$Ni$_{x}$)$_{2}X_{2}$ ($X=$ Se and S) has a unique mechanism, motivating us to explore the magnetic and electronic properties of Rb(Co$_{1-x}$Ni$_{x}$)$_{2}$Se$_{2}$ single crystals.

In this work, we report detailed studies on the crystallographic, magnetic structure, magnetic susceptibility, and electronic transport properties of Rb(Co$_{1-x}$Ni$_{x}$)$_{2}$Se$_{2}$ single crystals by utilizing powder x-ray diffraction (XRD), neutron powder diffraction (NPD), magnetic susceptibility, and resistance measurements. We find by increasing the Ni concentration, (1) the FM order transforms to AFM order at around $x=0.06$, (2) the AFM order extends to the compositions with $x=0.8$, (3) the anisotropic in-plane and out-of-plane magnetic susceptibilities reverse magnitudes at $x\approx0.3$, and (4) the $x=0.9$ compound is paramagnetic. 
Neutron diffraction measurements reveal that the AFM phase corresponds to a universal $A$-type magnetic structure with spins aligned in the $ab$ plane. While the moment size is increased from 0.69(2) for $x=0.1$ to 2.80(6)$\mu_B$ for $x=0.6$. The residual resistivity ratio ($RRR$) shows an intimate relationship with magnetism. The results suggest that the dilution of magnetic Co by nonmagnetic Ni has increased the localization of Co 3$d$ electrons. The magnetism evolves from itinerant to localized in the AFM region of the Rb(Co$_{1-x}$Ni$_{x}$)$_{2}$Se$_{2}$ phase diagram. Spin glass state is also identified in the compounds we explored. 

\begin{figure}[t]
	\centering
	\includegraphics[width=8cm]{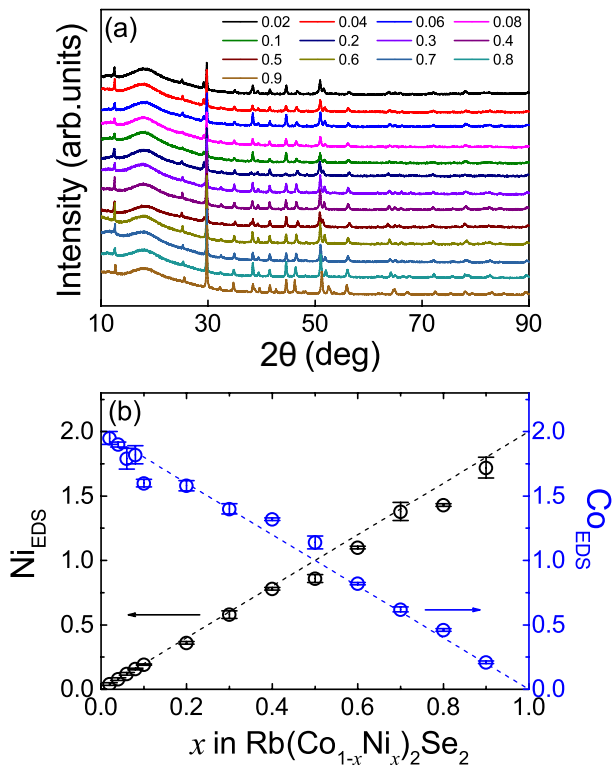}
	\caption{(a) XRD patterns of Rb(Co$_{1-x}$Ni$_{x}$)$_{2}$Se$_{2}$ powder samples with $0.02\leq x \leq0.9$. (b) EDS determined concentrations of Ni and Co as a function of the nominal concentration of Ni, $x$. Error bars are the uncertainty of the EDS results.}
	\label{fig1}
\end{figure}

\section{EXPERIMENTAL DETAILS}

Single crystals of Rb(Co$_{1-x}$Ni$_{x}$)$_{2}$Se$_{2}$ with compositions of $x$ = 0.02, 0.04, 0.06, 0.08, 0.1, 0.2, 0.3, 0.4, 0.5, 0.6, 0.7, 0.8, and 0.9 were synthesized by the Bridgman method with two-step reacting. 
Co$_{1-x}$Ni$_{x}$Se precursors were prepared by using Ni, Co powders, and Se shots to react at $500^\circ$C in an evacuated quartz tube. The process was repeated two times for homogeneity. Then, stoichiometric Rb ingot and Co$_{1-x}$Ni$_{x}$Se powders were placed in an aluminum crucible and sealed in an evacuated quartz tube. The mixtures were heated to $1070^\circ$C and kept for 10 h. 
After the samples were slowly cooled down to $700^\circ$C at a rate of $3.7^\circ$C/$h$, the furnace was shut down.\cite{Liu2022} 
We obtained plate-like single crystals with a typical size of $5\times5 \times1$ mm$^{3}$. 
As the Ni content increases, the samples have a metallic luster with colors from golden to pink. 
All samples are sensitive to oxygen and moisture. They were saved and handled in an argon-filled glove box. 

Powder XRD measurements were conducted on an x-ray diffractometer (Empyrean) at 300 K. NPD experiment was carried out on the multi-physics instrument (MPI) installed at the China Spallation Neutron Source (CSNS). 
Powder samples ground from single crystals of Rb(Co$_{1-x}$Ni$_{x}$)$_{2}$Se$_{2}$ with $x$ = 0.1, 0.4, and 0.6 were sealed in cylindrical vanadium cans with helium atmosphere and measured at 5, 50, and 200 K. NPD data were refined by employing the Rietveld method using the FULLPROF Suite software.\cite{Rietveld1969,Rodriguez-Carvajal1993}
The composition analysis was conducted on an energy dispersive x-ray spectroscopy (EDS) (EVO, Zeiss). 
Resistance, dc, and ac susceptibility were measured on a commercial physical property measurement system (PPMS, Quantum Design). 
The in-plane resistivity measurements were performed using the standard four-probe method.

\begin{figure}[t]
	\centering
	\includegraphics[width=7cm]{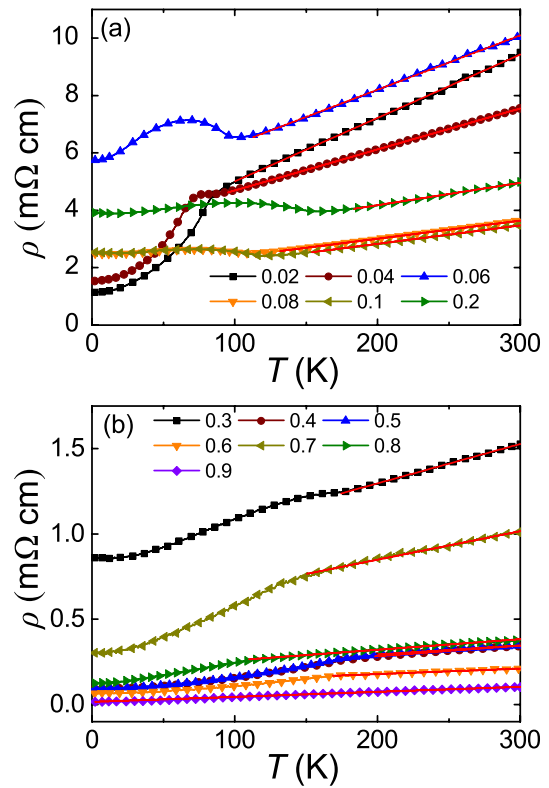}
	\caption{Temperature dependence of the in-plane resistivity $\rho_{ab}(T)$ for Rb(Co$_{1-x}$Ni$_{x}$)$_{2}$Se$_{2}$ with (a) 0.02 $\leq x \leq$ 0.2 and (b) 0.3 $\leq x \leq$ 0.9. The red-solid lines are linear fits to the resistivity above the magnetic transition temperature.}
	\label{fig5}
\end{figure}

\section{RESULTS AND DISCUSSION }

\begin{figure*}[t]
	\centering
	\includegraphics[width=17cm]{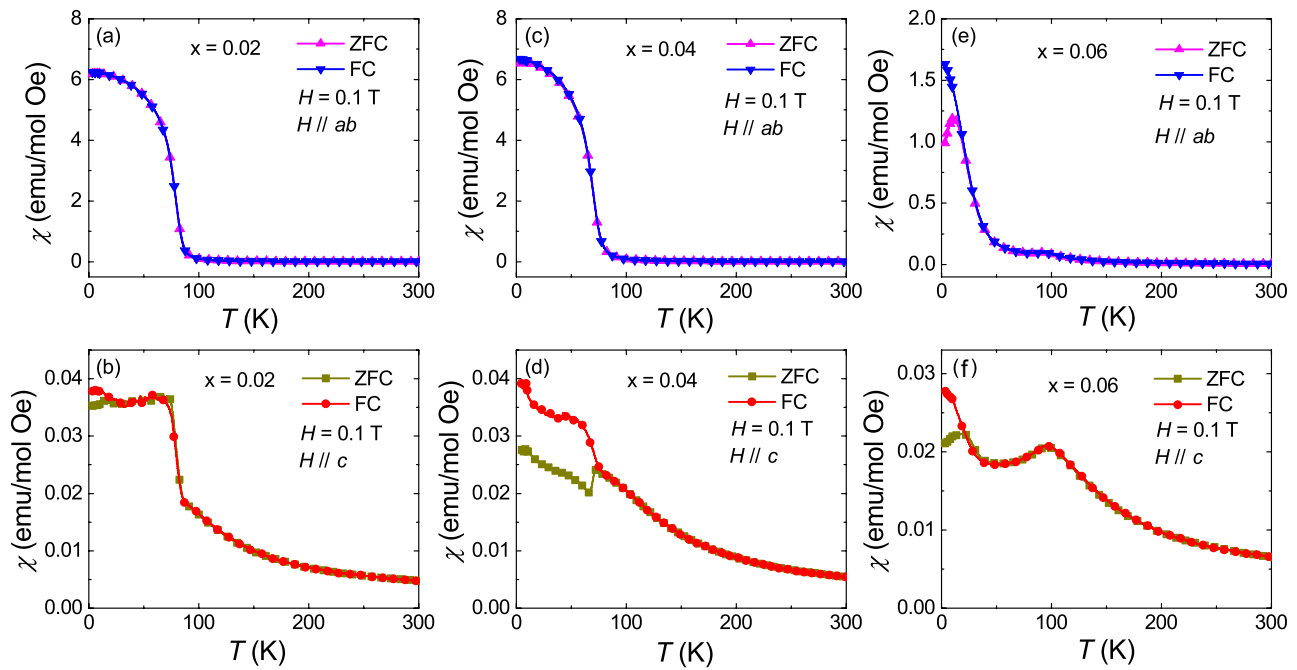}
	\caption{Temperature dependence of the magnetic susceptibilities $\chi(T)$ for Rb(Co$_{1-x}$Ni$_{x}$)$_{2}$Se$_{2}$ ($ x $ = 0.02, 0.04, and 0.06), measured in ZFC and FC conditions with \textit{H} = 0.1 T. The magnetic fields are parallel to the $ab$ plane in (a), (c), and (e), and parallel to the $c$ axis in (b), (d), and (f).}	
	\label{fig3}
\end{figure*}
	
Figure \ref{fig1}(a) shows powder XRD patterns at room temperature for the compositions of Rb(Co$_{1-x}$Ni$_{x}$)$_{2}$Se$_{2}$ (0.02 $\leq x \leq$ 0.9). The XRD patterns can be fitted well by the tetragonal ThCr$_{2}$Si$_{2}$-type structure with the space group $I4/mmm$. The fitted lattice constants for different compositions do not show a regular change, consistent with the expectation for the approximate ionic radii of Co and Ni. 
The average lattice parameters $a$ and $c$ from XRD measurements are 3.902(3) and 14.12(2) \AA, respectively. 
The actual compositions determined by the EDS for single crystals are plotted in Fig. \ref{fig1}(b). The content of Se was normalized to be 2 in the analysis. The linear relationships between the nominal and actual contents of Ni and Co suggest that the substitution is successful for all compositions. 

The resistivity [$\rho_{ab}$($T$)] measured in the $ab$-plane on single crystals of Rb(Co$_{1-x}$Ni$_{x}$)$_{2}$Se$_{2}$ in the temperature range from 2 to 300 K is shown in Fig. \ref{fig5}. All of the samples display metallic behavior. 
The resistivity decreases linearly as decreasing temperature before a drop appears at 86 and 72 K for the compounds with $x=0.02$ and 0.04, respectively [Fig. \ref{fig5}(a)]. The drop in resistivity could be attributed to a FM transition, as observed in FM RbCo$_{2}$Se$_{2}$ and Tl$_{1-x}$K$_{x}$Co$_{2}$Se$_{2}$, which can be understood by suppression of the magnetic scattering on conduction electrons in the FM state.\cite{Metals1989,Huang2021}
The anomalies can also be observed in the resistivity of Rb(Co$_{1-x}$Ni$_{x}$)$_{2}$Se$_{2}$ with $x=$ 0.06, 0.08, 0.1, 0.2, and 0.3 at higher temperatures. However, the resistivity shows a hump below the magnetic transition reminiscent of the behavior in CsCo$_{2}$Se$_{2}$, which undergoes an AFM transition at low temperatures.\cite{Yang2019b} Thus, the Ni substitution on the Co site may induce a FM to AFM ground state transition.
The Ni ions exhibit a paramagnetic characteristic in RbNi$_2$Se$_2$.\cite{Liu2022}
Further increasing the Ni content, the magnetic scattering would be reduced due to the decrease of the magnetic Co ions. The anomalies in resistivity for the compounds with $0.4\leq x\leq0.8$ indeed become weaker, as shown in Fig. \ref{fig5}(b). For the composition with $x$ = 0.9, the $\rho_{ab}$($T$) does not show any anomaly in the measured temperature range, similar to RbNi$_2$Se$_2$.

\begin{figure*}[t]
	\centering
	\includegraphics[width=17cm]{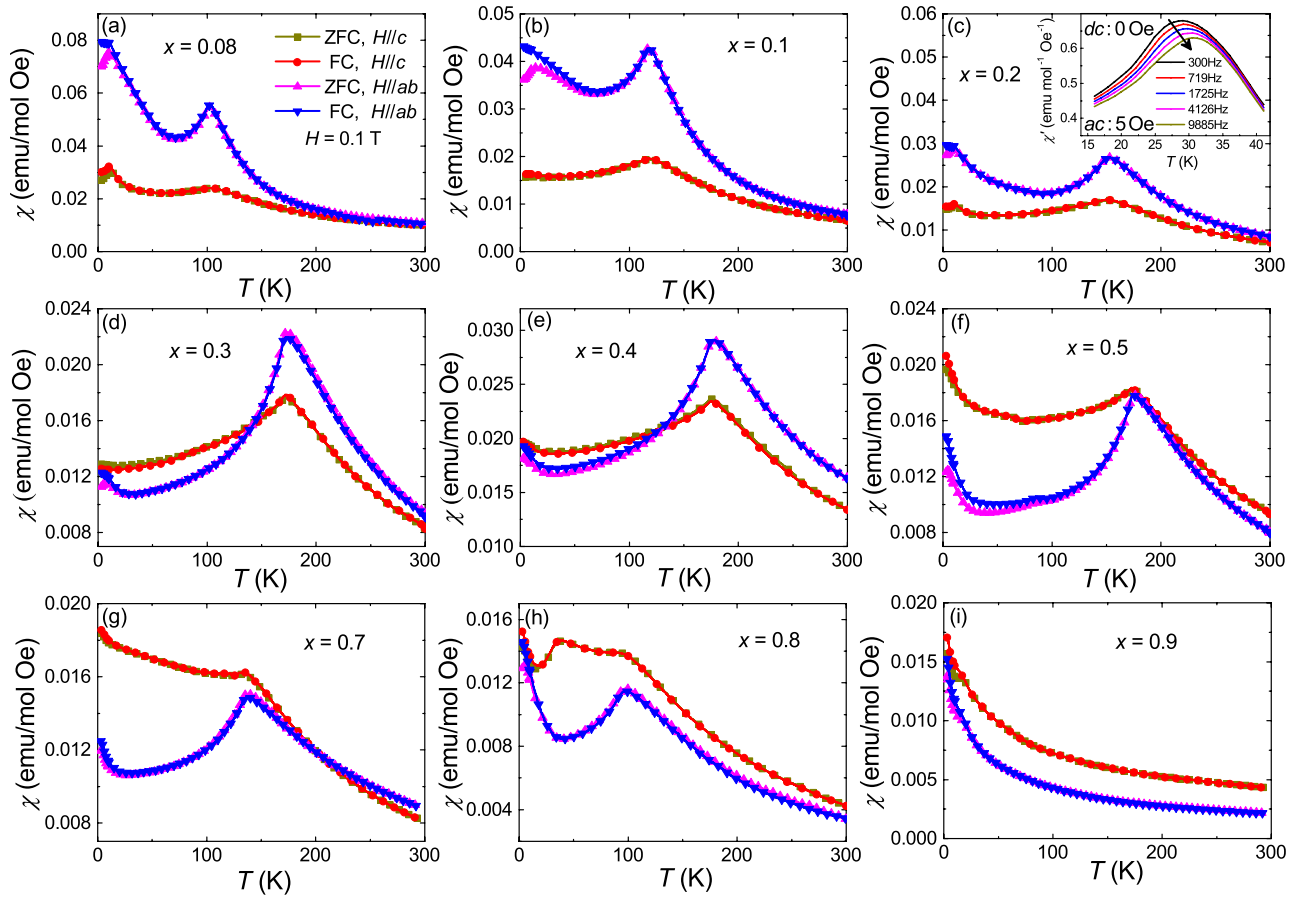}
	\caption{Temperature dependence of $\chi(T)$ measured in ZFC and FC conditions for \textit{H} $\|$ \textit{ab} and \textit{H} $\|$ \textit{c} with \textit{H} = 0.1 T for Rb(Co$_{1-x}$Ni$_{x}$)$_{2}$Se$_{2}$ with (a) $x=0.08$, (b) $x=0.1$, (c) $x=0.2$, (d) $x=0.3$, (e) $x=0.4$, (f) $x=0.5$, (g) $x=0.7$, (h) $x=0.8$, and (i) $x=0.9$. The Inset in (c) is the temperature dependence of the $\chi'(T)$ measured at various $ac$ magnetic field frequencies between 300 and 9885 Hz for the compound with $x$ = 0.20. The $ac$ magnetic field is parallel to the $ab$ plane. The arrow indicates a shift in the freezing temperatures.}	
	\label{fig4}
\end{figure*}

To investigate the magnetic properties, we measured the magnetic susceptibilities [$\chi$($T$)] for all the compounds, as shown in Figs. \ref{fig3} and \ref{fig4} .
Figure \ref{fig3} shows the $\chi$($T$) under zero-field-cooled (ZFC) and field-cooled (FC) conditions for the Rb(Co$_{1-x}$Ni$_{x}$)$_{2}$Se$_{2}$ crystals with $x$ = 0.02, 0.04, and 0.06.
The susceptibility data were collected with the magnetic field applied along the $ab$ plane [$\chi_{ab}$] and the $c$ axis [$\chi_{c}$], respectively. The abrupt enhancement of $\chi_{ab}$ below 100 K reveals the development of a long-range FM order, as shown in Fig. \ref{fig3}(a), (c), and (e) for the compounds $x$ = 0.02, 0.04, and 0.06. The enhancement could also be observed on the $\chi_{c}$. However, the strength of the magnetic response within the layer [$\chi_{ab}$] shows an order of magnitude larger than that of the out-of-layer [$\chi_{c}$], yielding that the ferromagnetically ordering moments are aligned in the $ab$ plane. 
For the compound with $x$ = 0.06, the out-of-plane susceptibility displays a cusp-like peak around 96 K [Fig. \ref{fig3}(f)], in contrast to the in-plane FM behavior, implying that an AFM order may occur along the $c$ axis. 

The magnetic susceptibilities for $x$ = 0.08, 0.1, 0.2, 0.3, 0.4, 0.5, 0.7, 0.8, and 0.9 are shown in Figs. \ref{fig4}(a)-\ref{fig4}(i) with $H\|ab$ and $H\|c$, respectively. The cusp-like anomalies are more obvious in the susceptibilities of Rb(Co$_{1-x}$Ni$_{x}$)$_{2}$Se$_{2}$ with 0.08 $\leq x \leq$ 0.8. The magnitudes of $\chi_{ab}$ and $\chi_{c}$ are comparable and consistent with an AFM order. Interestingly, there is an obvious magnitude change between the $\chi_{ab}$ and $\chi_{c}$ in the composition range of $x=0.08$ to $x=0.8$, where the magnitude reverses at $x=0.3$. It is difficult to determine the origin of the changes in magnetic susceptibility because of the unknown AFM order and the substitution of the magnetic Co ions.

Comparing the ZFC to FC magnetic susceptibilities, a pronounced bifurcation can be observed below a transition temperature for the compositions with 0.06 $\leq x \leq$ 0.9.
The irreversible behavior implies a spin-glass (SG) state, which may be related to the doping of the nonmagnetic Ni ions. 
The frequency-dependent $ac$ susceptibility measurements were adopted to verify the SG state for the compound with $x$ = 0.2. The magnetic field was applied parallel to the $ab$ plane. As shown in the inset of Fig. \ref{fig4}(c), the real part of the $ac$ susceptibilities $\chi'$($T$) exhibit a clear frequency dependence. With increasing frequency, the peak position shifts to higher temperatures while the peak amplitude decreases.
The Mydosh parameter ($\delta T_{f}$) represents the relative shift in freezing temperature per decade of frequency, which can be calculated by $\delta T_{f}$ = $\Delta T_{f}$ / ($T_{f}$ $\Delta$log$f$).\cite{Ryu2015} The spin freezing temperature $T_{f}$ is defined by the peak position from the ZFC curve.
The obtained $\delta T_{f}$ value is 0.031, consistent with the values of 0.0045 $\leq$ $\delta T_{f}$ $\leq$ 0.08 for a typical SG state. For the compound with $x$ = 0.9 in Fig. \ref{fig4}(i), the magnetic transition disappears in the temperature dependence of susceptibility. The $\chi(T)$ curves exhibit the Pauli PM behavior without long-range magnetic order, similar to the characteristic of RbNi$_{2}$Se$_{2}$.\cite{Liu2022}

\begin{figure}[t]
	\centering
	\includegraphics[width=7cm]{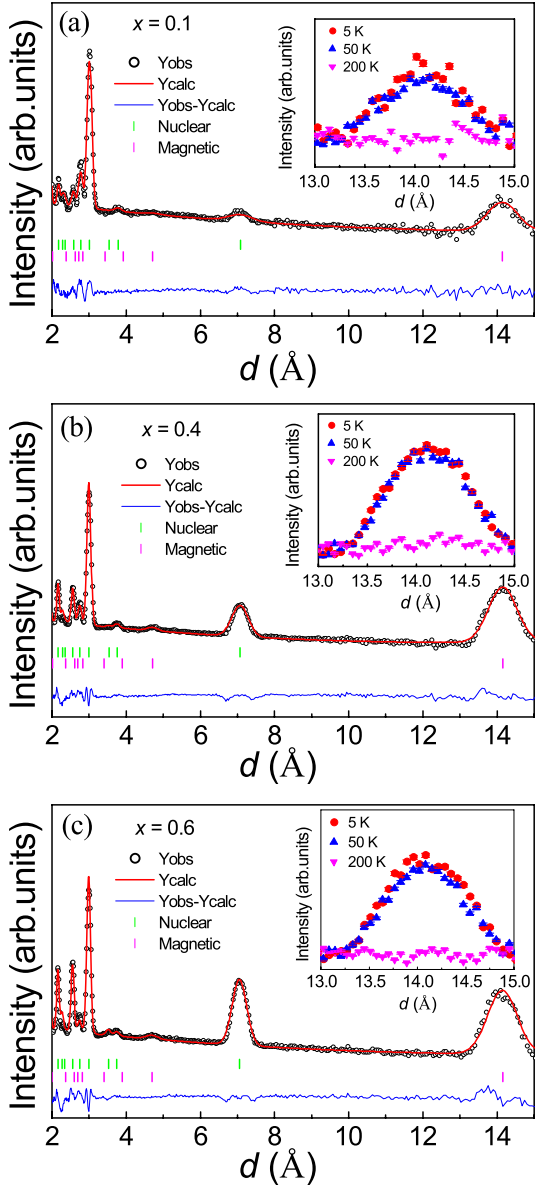}
	\caption{NPD patterns for Rb(Co$_{1-x}$Ni$_{x}$)$_{2}$Se$_{2}$ at 5 K with (a) $x$ = 0.1, (b) $x$ = 0.4, and (c) $x$ = 0.6. The insets are temperature dependence of the magnetic diffraction peak corresponding to the (001) reflection at 5, 50, and 200 K. The errors in the insets indicate one standard deviation. }
	\label{fig2}
\end{figure}

\begin{figure}[t]
	\centering
	\includegraphics[width=8cm]{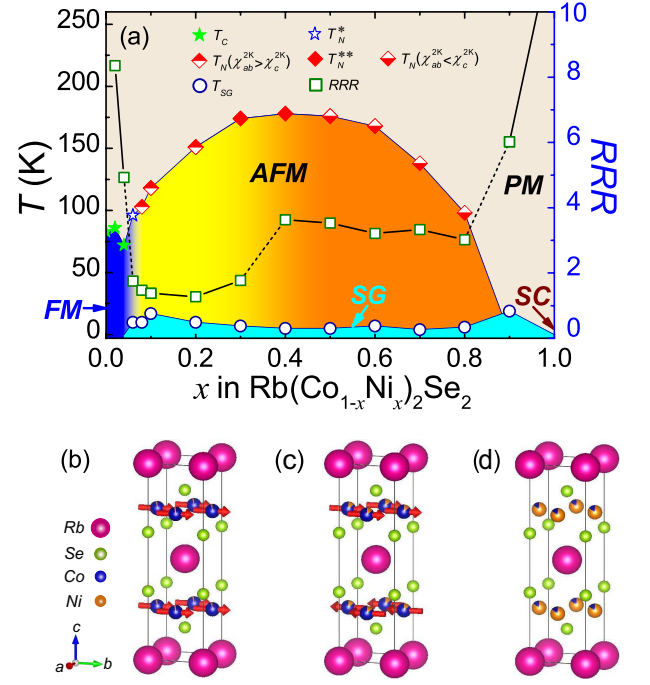}
	\caption{(a) A phase diagram of Rb(Co$_{1-x}$Ni$_{x}$)$_{2}$Se$_{2}$ with the FM transition temperature [$T_{\text{C}}$], AFM transition temperature [$T_{\text{N}}$, $T_{\text{N}}^{\ast}$, $T_{\text{N}}^{\ast\ast}$], spin glass transition temperature [$T_{\text{SG}}$], and the residual resistivity ratio [$RRR$].  $T_{\text{N}}^{\ast}$ represents the AFM transition temperature for the critical concentration $x$ = 0.6 with FM behavior on $\chi_{ab}$ and AFM behavior on $\chi_{c}$. $T_{\text{N}}^{\ast\ast}$ is the AFM transition temperature for $x$ = 0.3 and 0.4, where the anisotropic susceptibility curve has a crossover below $T_{\text{N}}$. Schematic diagrams of (b) the FM structure, (c) the AFM structure, and (d) the PM structure in the phase diagram.}	
	\label{fig6}
\end{figure}

To obtain accurate magnetic structures, NPD data were collected for samples with $x$ = 0.1, 0.4, and 0.6. The reflection patterns and the Rietveld method refinements are shown in Fig. \ref{fig2}. By comparing the data at 5, 50, and 200 K, a magnetic peak at $d=14.2$ \AA\ can be identified.
The magnetic peak can be indexed with the propagation vector $\textbf{\textit{k}}$ = $[1,0,0]$ in the notation of the $I4/mmm$ space group with $A$-type AFM order, as shown in Fig. \ref{fig6}(c). The magnetic diffraction peak observed at $d$ = 14.2 \AA\ corresponds to the magnetic Bragg reflection $(H, K,L) = (0 0 1)$. Here, we assume the magnetism arising from the Co ions.
The ordered moments are refined to be 0.69(2), 1.98(4), and 2.80(6) $\mu_\text{{B}}$ per Co at 5 K for the compounds with $x$ = 0.1, 0.4, and 0.6, respectively. The increased ordered moment for the $x=0.6$ compound is close to the fully localized high spin state of Co$^{2+}$ with $S=3/2$.

Knowing the $A$-type magnetic order and moment sizes for the compounds with 0.08 $\leq x \leq$ 0.8, the evolution of the magnetic susceptibilities can be understood. By increasing the content of Ni, the magnetism evolves from itinerant to localized, accompanied by the enhancement of the ordered moment size. The AFM intralayer coupling between the FM layers will be increased accordingly.
 For the compounds with $x$ = 0.08, 0.1, and 0.2, the $\chi_{ab}$ sharply drops at the magnetic transition temperature $T_{\text{N}}$. The $\chi_{c}$ has a weak temperature dependence below $T_{\text{N}}$, typically conforming the $A$-type AFM structure with the magnetic moments lying along the $ab$-plane. The value of in-plane susceptibility $\chi_{ab}$ is enhanced by the $ab$-plane collinear ferromagnetic layers, larger than the out-of-plane susceptibility $\chi_{c}$. A similar magnetic structure has been reported in CsCo$_{2}$Se$_{2}$, which exhibits the approximate magnetic behavior.\cite{Yang2013,Yang2019b}
When $x$ = 0.3 and 0.4, the susceptibilities exhibit a crossover between the $\chi_{ab}$ and $\chi_{c}$.
Such a crossover could be ascribed to the dilution of the magnetic Co ions in the $ab$ plane and the enhancement of the AFM intralayer coupling.

By combining the experimental results on Rb(Co$_{1-x}$Ni$_{x}$)$_{2}$Se$_{2}$, we construct a comprehensive phase diagram in Fig. \ref{fig6}(a), summarizing the evolutions of the magnetic order, spin glass state, and $RRR$ as a function of the Ni content $x$.
For $x$ = 0.02 and 0.04, the anisotropic magnetic susceptibilities reveal a FM order with moments aligned in the $ab$ plane, as shown in Fig. \ref{fig6}(b). Both the FM and AFM transitions could be identified in the compound with $x$ = 0.06, suggesting that the FM to AFM transformation is a first-order transition and $x$ = 0.06 is close to the critical composition. The AFM structure is shown in Fig. \ref{fig6}(c). For the compounds with AFM order, the magnetic susceptibility changes from $\chi_{ab}>\chi_{c}$ to $\chi_{ab}<\chi_{c}$ at the compositions with $x=$ 0.3 and 0.4, resulting from the dilution of the magnetic Co ions and localization of the magnetic moments. The AFM transition vanishes from the susceptibility measurements for the compound with $x$ = 0.9 [Fig. \ref{fig6}(d)]. The SG transition temperature [$T_{\text{SG}}$] is defined as the bifurcation point between the ZFC and FC susceptibilities. The $T_{\text{SG}}$ and $RRR$ for each composition are plotted in Fig. \ref{fig6}(a). 
The evolution of the electronic behavior ($RRR$) is strongly correlated with the magnetic order. The abrupt increase of $RRR$ in the AFM region could be attributed to the enhancement of the localization of Co ions. To further understand the nature of magnetic and transport properties, angle-resolved photoemission spectroscopy studies on the electronic band structures are required.

\section{SUMMARY}

In conclusion, we have grown the single crystals of Rb(Co$_{1-x}$Ni$_{x}$)$_{2}$Se$_{2}$ with 0.02 $\leq x \leq$ 0.9 and comprehensively investigated the physical properties via the XRD, EDS, NPD, magnetic susceptibility, and electronic transport methods. By combining these results, we present a phase diagram of Rb(Co$_{1-x}$Ni$_{x}$)$_{2}$Se$_{2}$. 
All compositions adopt the space group $I4/mmm$ and show a slight difference in lattice parameters.
The magnetism evolves from FM to AFM, then to Pauli paramagnetism with increasing the concentration of Ni. The electronic transport behaviors are closely correlated with the magnetic order.
The magnetic structure in the AFM region is confirmed to be the $A$-type AFM order by NPD. The AFM moment size is increased significantly as increasing the Ni concentration, demonstrating that the magnetic moments of Co ions become more localized as the dilution effect by the Ni substitution.

\section{ACKNOWLEDGMENTS}

Work was supported by the Guangdong Basic and Applied Basic Research Foundation (Grants No. 2021B1515120015, 2020B151520065), the National Natural Science Foundation of China (Grant No. 12174454), the Guangzhou Basic and Applied Basic Research Foundation (Grant No. 202201011123), Guangdong Provincial Key Laboratory of Magnetoelectric Physics and Devices (Grant No. 2022B1212010008), and National Key Research and Development Program of China (Grant No. 2019YFA0705702).
	
\bibliography{reference}

\end{document}